\documentclass{roajarticle}
  \usepackage{graphicx}
  \usepackage{graphics}
  \usepackage[section]{placeins}
  \usepackage{natbib}
  \usepackage{amsmath}
  \usepackage{multirow}
  \usepackage[table]{xcolor}
  \usepackage{grffile}
  \usepackage[nodisplayskipstretch]{setspace}
  \usepackage{float}
  \usepackage{upgreek}
  \usepackage{epstopdf}
  \usepackage{rotating}
  \usepackage[flushleft]{threeparttable}
  
  \usepackage{xcolor}
  
  \newcommand{\volume}{\textbf{...}}

 \title{Statistical analysis of astro-geodetic data through principal component analysis, linear modelling and bootstrap based inference}
  \author[1,2]{Andreea Ioana \MakeTextUppercase{Gornea}}
  \author[1]{Alexandru \MakeTextUppercase{Calin}}
  \author[1]{Paul Daniel \MakeTextUppercase{Dumitru}}
  \author[2]{Dan Alin \MakeTextUppercase{Nedelcu}}
   \author[3]{Radu Stefan  \MakeTextUppercase{Stoica}}

  \affil[1]{Technical University of Civil Engineering Bucharest\\%
  Lacul Tei Bvd. 122 - 124, 020396 Bucharest, Romania,
   gornea.andreea@gmail.com}

  \affil[2]{Astronomical Institute of Romanian Academy\\%
   Str. Cutitul de Argint 5, 040557 Bucharest, Romania} 

\affil[3]{Universit\'e de Lorraine, Institut Elie Cartan de Lorraine\\%
54506 Vandoeuvre-l\'es-Nancy Cedex, France}
  

\begin{document}
  
\maketitle

\begin{abstract}
The paper demonstrates the application of statistical based methodology for the analysis of the vertical deviation angle. The studied data set contains astro-geodetic observations. The Principal Component Analysis and the Multiple Linear Regression models are embedded within a bootstrap procedure, in order to overcome the difficulties related to data correlation, while taking advantage of all the information provided. The methodology is applied on real data. The obtained results indicate that the pressure, the temperature and the humidity are variables that may influence the measure of the vertical deviation.
\end{abstract}

\section{Introduction}

The vertical deviation angle is a notion of great interest in the field of Geodesy, since it is used to establish a link between two surfaces that approximate the shape of the Earth: the geoid and the ellipsoid. The vertical deviation is defined by the angle between the direction of the plumb line and the normal to the ellipsoid through the same point on the surface of Earth \citep{Featherstone1999}. 

The vertical deviation angle can be obtained with a geodetic total station that has attached a CCD camera and GNSS (Global Navigation Satellite System)  receiver. This type of ensemble measures the azimuth, the zenith distance of a star that crosses its reticular wires and the time of these intersections. The horizontal coordinates are transformed into astronomic coordinates ($\Phi$, $\Lambda$) of the observation location. The geodetic coordinates ($\varphi$, $\lambda $)  of the observation location are obtained with a GNSS instrument (which has as reference surface an ellipsoid). The vertical deviation orthogonal components are obtained with the equation (\ref{eqxieta}), where $\xi$ is the meridian component - on North South direction and $\eta$ is the prime vertical component - on East West direction.
	
\begin{equation}
\xi = \Phi - \varphi \quad
;
\quad \eta = (\Lambda - \lambda )cos\varphi 
\label{eqxieta}
\end{equation}

The data set analysed in the paper is a large table containing the vertical deviations and its associate measures. These data were recorded on the concrete pilaster from the roof of the Faculty of Geodesy within the Technical University of Civil Engineering Bucharest (FG-TUCEB) on twelve nights.  The data structure is given in Table~\ref{dataset}. To each deviation angle nine quantitative variables are attached. Columns 2, 3 and 4 contain the atmospheric parameters, columns 5 and 7 contain errors given by the CCD image acquisition software. These values represent the errors resulted after the trajectory of the star in the field of view was reconstructed based on all the CCD images. Columns 8, 9 and 10 are variables that characterise the observed star. The data contained in Table~\ref{dataset} is not normalised. \\

\begin{table}\label{dataset}
\centering
  \begin{threeparttable}
{\footnotesize
\caption{Data set} 
\begin{tabular}[H]{|c|c|c|c|c|c|c|c|c|c|}
\hline
	Star	& P & T &H&rms1&img&rms2&A&z&V\\ 
&(HPa)&($^{\circ}{\rm C}$)&(\%)&(px)&&(px)&($^{\circ}...$)&($^{\circ}...$)&($^{\circ}$/h)\\ \hline
\color{gray}1&\color{gray}2&\color{gray}3&\color{gray}4&\color{gray}5&\color{gray}6&\color{gray}7&\color{gray}8&\color{gray}9&\color{gray}10\\ \hline
$\alpha Cas$&1002.1&8.6&50.0&0.15&161&0.2&322.28286&60.91284	&8.27\\ \hline 
$\alpha Ori$	&1002.3&8.0&50.0&0.21&312&0.16&231.03903&48.43744&14.87 \\ \hline 	
$\gamma Cep$& 1002.3&7.4&50.0	&0.21&263&0.23&350.05359&55.17984& 3.21\\ \hline
$\alpha Hya$& 1002.4&7.1&50.0	&0.24&523&0.19&176.01978&53.27512&14.82\\ \hline
\end{tabular}}
\begin{tablenotes}
      \small
      \item 1 - Observed star, 2 - Pressure, 3 - Temperature, 4 - Humidity, 5 -root mean square 1 in pixels, 6 - number of images taken by the CCD camera, 7 - root mean square 2 in pixels, 8 - Azimuth, 9 -  Zenith distance, 10 - Star apparent velocity 
    \end{tablenotes}
\end{threeparttable}
\end{table}	

\begin{figure}[H]
\centering
\includegraphics[scale=0.24]{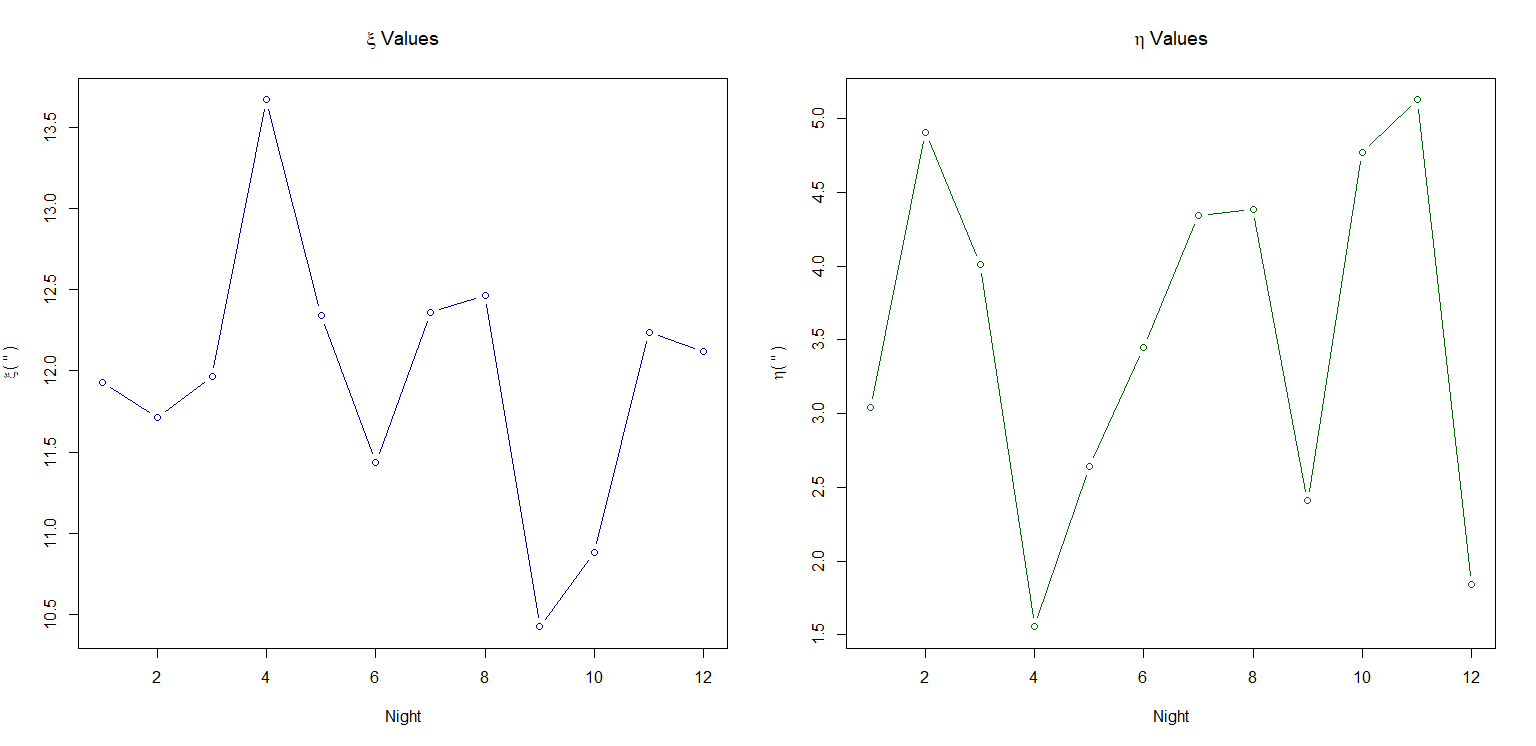}
\caption{$\xi$ and $\eta$ values variation. Values from all observation nights (12) - FG - TUCEB show a significant variation relative to the value size}
\label{xietavalues}
\end{figure}

Figure~\ref{xietavalues} shows the behaviour of the measured values $\xi$ and $\eta$ given the deviation angle. The present paper has two aims. The first one is to exhibit possible linear relations between the deviation angle values and their associate measures, respectively. The second one  attempts to establish which of these measures plays an important role in the variability of the deviation angle. The answer to these questions are formulated proposing a bootstrap based statistical analysis based on Principal Component Analysis (PCA) and Multi-Linear Regression (MLR). The vertical deviation components behaviour resembles the one of a statistical series. Within this assumption, the present paper aims to study the variability of the data set that contains the vertical deviation components. 

The structure of the paper is as follows. The next two sections present the PCA and MLR methods and results applied to our data set, together with their associate bootstrap procedure respectively. Finally, conclusions and perspectives are depicted.

\section{PRINCIPAL COMPONENT ANALYSIS}
Principal Component Analysis (PCA) is a statistical method allowing numerical and visual description of relevant features in multidimensional data~\citep{Husson2011}. The data set from the present study has 510 rows (individuals = measurements) and 9 columns (quantitative variables). Additionally, the data set contains two supplementary quantitative variables - the two components of the vertical deviation angle and two qualitative variables -  the observation night and the observed star.

 PCA was applied on the standardised data set because the variables did not have the same measurement unit. Standardisation means considering $\frac{x_{ik}-\bar{x_k}}{s_k}$ instead of $x_{ik}$ and choosing standard deviation $s_k = \sqrt{\frac{1}{I-1}\sum\limits_{i=1}^I(x_{ik} - \bar{x_k})^2}$ as a unit of measurement for each variable $k$. Connecting these new values with the normal distribution can highlight possible outliers or extreme values. Table \ref{exstd} contains a few standardised values for all nine quantitative variables. The measures higher than $2$ in absolute value are coloured in blue. Assuming a Gaussian character for the observed data, the following variables may be considered rather extreme: humidity, number of images, root mean square $2$ and velocity.
 
\begin{table}
\centering
{\footnotesize
\caption{Suggested extreme values in standardised data set} \label{exstd}
\begin{tabular}[H]{|c|c|c|c|c|c|c|c|c|c|c|}
\hline
Ind &Star	& P & Temp &Hum&rms1&No img&rms2&A&z&V\\ \hline
\color{gray}1&\color{gray}2&\color{gray}3&\color{gray}4&\color{gray}5&\color{gray}6&\color{gray}7&\color{gray}8&\color{gray}9&\color{gray}10&\color{gray}11\\ \hline
56&$\alpha Hya$&0.30&-1.85&-0.13&-0.44&\cellcolor{blue!15}3.73&-0.89&0.04&-0.44&0.81\\ \hline
57&$\beta UMi$&-1.94&-1.62&-0.13&1.95&1.85&\cellcolor{blue!15}4.14&-1.47&-0.79&-1.93\\ \hline
58&$\beta UMi$&-1.83&-1.64&-0.13&-0.44&\cellcolor{blue!15}3.37&0.52&-1.47&-0.82&-1.93\\ \hline
59&$\beta UMi$&-1.85&-1.64&-0.13&0.04&\cellcolor{blue!15}2.12&0.52&-1.47&-0.84&-1.93\\ \hline
60&$\beta UMi$&-1.85&-1.65&-0.13&1.00&1.86&\cellcolor{blue!15}2.57&-1.47&-0.85&-1.93\\ \hline
61&$\beta UMi$&-1.83&-1.68&-0.13&0.52&\cellcolor{blue!15}4.88&1.15&-1.47&-0.88&-1.93\\ \hline
62&$\beta UMi$&-1.81&-1.69&-0.13&-0.44&\cellcolor{blue!15}2.89&0.52&-1.47&-0.90&-1.93\\ \hline
63&$\beta UMi$&-1.81&-1.71&-0.13&-0.28&\cellcolor{blue!15}2.12&0.68&-1.47&-0.92&-1.93\\ \hline
64&$\beta UMi$&-1.81&-1.71&-0.13&0.20&\cellcolor{blue!15}2.42&\cellcolor{blue!15}2.57&-1.47&-0.93&-1.93\\ \hline
506&$\eta UMa$&1.96&-0.52&\cellcolor{blue!15}3.73&1.00&0.00&0.37&1.35&0.63&-0.48\\\hline
507&$\eta UMa$&1.96&-0.52&\cellcolor{blue!15}3.73&0.84&0.01&0.21&1.35&0.66&-0.48\\\hline
508&$\alpha UMi$&1.96&-0.55&\cellcolor{blue!15}3.94&0.20&-0.46&-0.11&1.83&-1.59&\cellcolor{blue!15}-2.93\\\hline
509&$\alpha UMi$&1.96&-0.58&\cellcolor{blue!15}4.14&1.00&0.20&0.21&-1.66&-1.60&\cellcolor{blue!15}-2.93\\\hline
510&$\alpha UMi$&1.96&-0.58&\cellcolor{blue!15}4.14&1.00&0.20&0.21&-1.66&-1.60&\cellcolor{blue!15}-2.93\\\hline
\end{tabular}}
\end{table}

PCA considers the data set to be analysed as a point cloud in a multidimensional space. Through PCA the best viewpoint is searched by finding the dimensions (axes) on which the variability of the point cloud is higher. Whenever the individuals are studied, the data set is seen as a point cloud in a space with $9$ dimensions (the number of active quantitative variables). Whenever the variables are studied the data set is considered as a point cloud in a space with $510$ dimensions (the number of individuals).

\begin{figure}[H]
\centering
\includegraphics[scale=0.5]{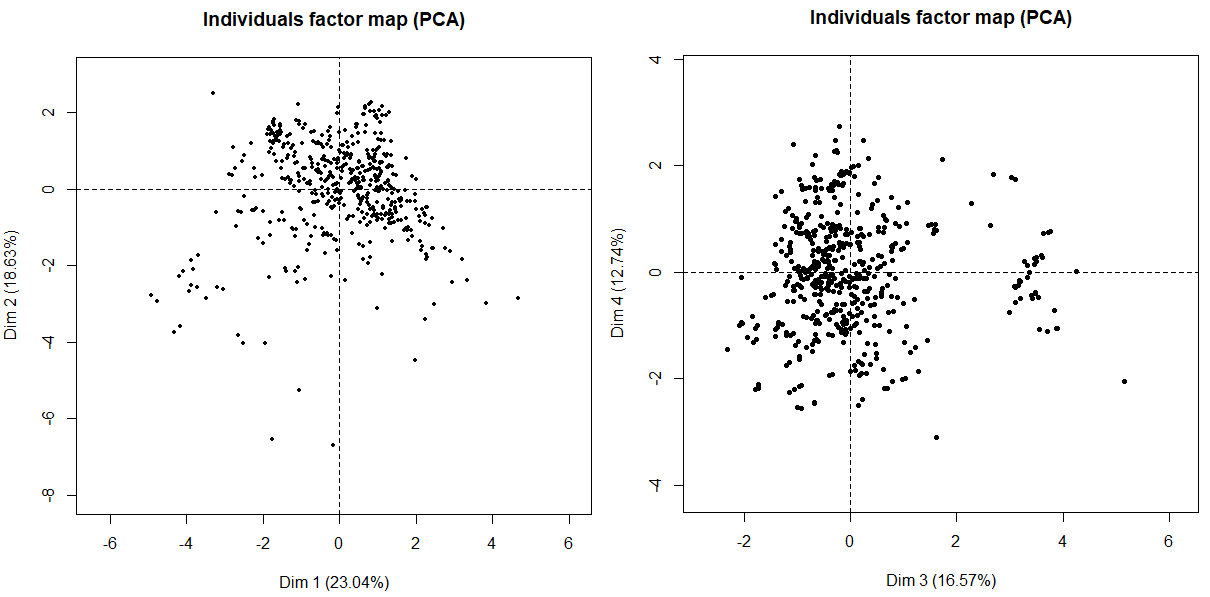}
\caption{Visualisation of the first four axes given by the PCA, $(D_i), i=1,\ldots,4$. Left plot : the data point cloud projected on the $(D_1,D_2)$ plane. Right plot : the data point cloud projected on the $(D_3,D_4)$ plane.}
\label{indfactmap}
\end{figure}

The Figure \ref{indfactmap} shows the projection of the data set point cloud on the axes given by the PCA analysis. It can be observed that the majority of points tend to be concentrated in a rather compact region around the origin, while the rest of them tend to spread further away from it within a certain degree of variability. The variability of a dimension is measured as the ratio of the inertia of the cloud projected on the considered dimension over the total inertia of the cloud. Correlating Table \ref{exstd} and Figure \ref{indfactmap} reveal which individuals can be considered as extreme or outliers. 

The variability percentages caught by each dimension, respectively, are presented in Table~\ref{variance}. The axes are usually, ordered and presented in decreasing order with respect to the variability. This PCA feature allows the reduction of the number of dimensions, since we may consider only those dimensions that have an important contribution to the total cloud inertia.

\begin{table}
\centering
\caption{Percentage of inertia for each dimension and cumulative percentage of inertia}
{\footnotesize
\begin{tabular}[!h]{|c|c|c|}
\hline
Dimension&Percentage of inertia&Cumulative percentage of inertia\\ \hline
1&$23.04\%$&$23.04\%$ \\ \hline
2&$18.62\%$&$41.66\%$\\ \hline
3&$16.56\%$&$58.23\%$\\ \hline
4&$12.74\%$&$70.98\%$\\ \hline
5&$9.53\%$&$80.51\%$\\ \hline
6&$7.25\%$&$87.77\%$\\ \hline
7&$5.36\%$&$93.14\%$\\ \hline
8&$3.74\%$&$96.89\%$\\ \hline
9&$3.11\%$&$100\%$\\ \hline
\end{tabular}
}
\label{variance}
\end{table}

\begin{figure}[H]
\centering
\includegraphics[scale=0.5]{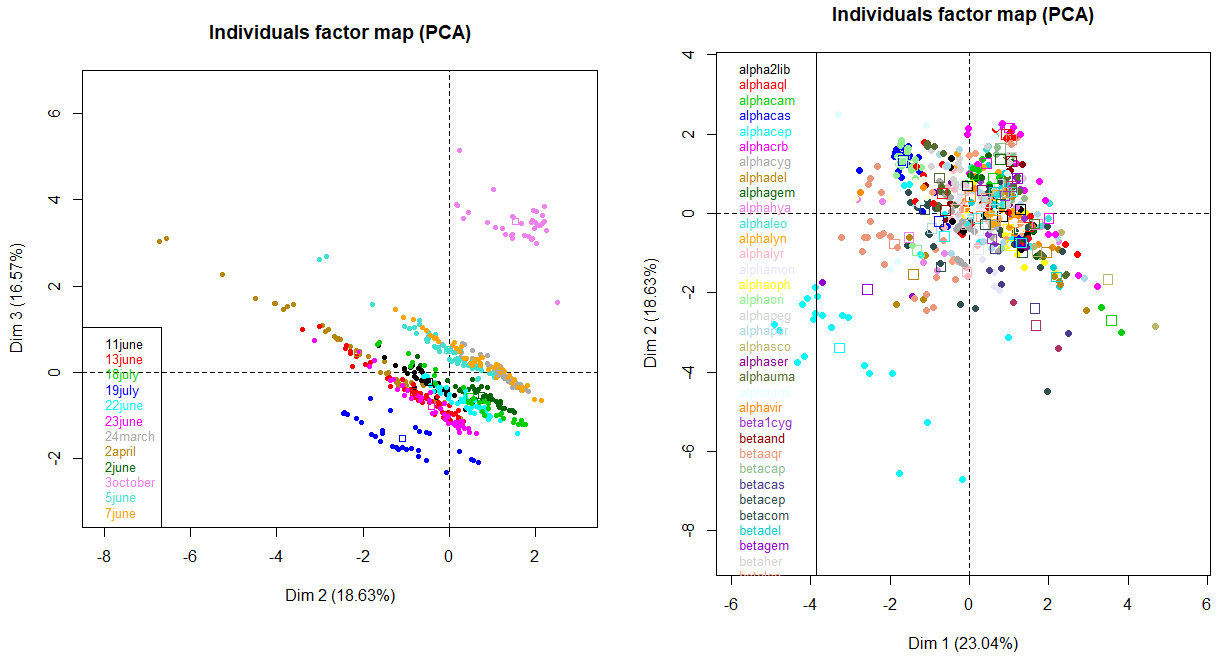}
\caption{Color coded projection of the data set point cloud. Left plot~: $(D_2, D_3)$ plane projection with the colours representing the observation nights. Right plot~: $(D_1, D_2)$ plane projection with the colours representing the observed star.}
\label{starnight}
\end{figure}

The PCA enables to see whether different measures tend to exhibit a certain pattern within the data set point cloud. The left plot in Figure~\ref{starnight} shows the point cloud projected on the $(D_2,D_3)$ plane, while the points are labelled depending on the night the measure was obtained. The right plot in Figure~\ref{starnight} shows the point cloud projected on the $(D_1,D_2)$ plane, while the points are labelled depending on the observed star. A clustered pattern depending on the observation is observed in the left plot. The pattern in the right plot appears to be less clustered, much more mixed, and it does not allow to affirm just by visual inspection how the choice of the observed star influences the distribution of points in the cloud.

It is legitimate to ask whether these patterns appeared by chance or not. In order to answer this question, a statistical test based on a bootstrap procedure is implemented. 

The test works as follows. First, for each cluster the median centre is computed. This choice is adopted against the most common one of the gravity centre, in order to use a measure much more robust to extreme values. Then the labels of the point cloud were changed uniformly, while keeping the point positions fixed. A number of $50$ such samples was obtained. For each such simulated data set the median centres were computed. This allows testing whether the observed median centres may be considered as the outcome of "completely random" distribution of labels. 

Tables \ref{mediancoord} contains the quantile interval, defined by the second highest median coordinate (column 4) and second lowest median coordinate (Column 3) of the rearranged groups of individuals defined by the observation night. Column 5 contains the initial medians and the ones that are highlighted are inside the quantile interval. If the initial median is outside the quantile interval, the test rejects the null hypothesis, that the clustered structure from Figure \ref{starnight} occurred by chance. As it can be observed in most of the cases, the medians are outside the intervals. Hence, in these situations, there is statistical evidence that the clusters induced by the observation do not occur simply by chance.

\begin{table}
\centering
{\footnotesize
\caption{Confidence intervals for each individual group defined by the observation night - dimensions 2 and 3} \label{mediancoord}
\begin{tabular}[H]{|c|c|c|c|c|}
\hline
Group/Date&Dimension&$Q_{50}^2$&$Q_{50}^{49}$& Initial median	\\ \hline
\color{gray}1&\color{gray}2&\color{gray}3&\color{gray}4&\color{gray}5\\ \hline
\multirow{2}{*}{24-Mar}&2&-0.2121&0.4235&1.2785 \\
&3&\cellcolor{red!15}\cellcolor{red!15}-0.4537&\cellcolor{red!15}-0.0761&\cellcolor{red!15}-0.1209 \\ \hline
\multirow{2}{*}{02-Apr}&2&-0.2441&0.5379&-1.6020\\
&3&-0.5381&-0.0122&0.2494\\ \hline
\multirow{2}{*}{02-Jun}&2&-0.2006&0.5055&0.8056\\
&3&-0.5244&0.0567&-0.5717\\ \hline
\multirow{2}{*}{05-Jun}&2&\cellcolor{red!15}-0.1800&\cellcolor{red!15}0.6150&\cellcolor{red!15}0.1404\\
&3&-0.4142&-0.0383&0.3114\\ \hline
\multirow{2}{*}{07-Jun}&2&-0.3461&0.4347&0.7221\\
&3&-0.4935&0.1442&0.2037\\ \hline
\multirow{2}{*}{11-Jun}&2&-0.2945&0.4752&-0.5074\\
&3&\cellcolor{red!15}-0.4429&\cellcolor{red!15}0.0039&\cellcolor{red!15}-0.2412\\ \hline
\multirow{2}{*}{13-Jun}&2&-0.2730&0.5352&-0.6396\\
&3&-0.4399&-0.0477&-0.5899\\ \hline
\multirow{2}{*}{22-Jun}&2&\cellcolor{red!15}-0.4111&\cellcolor{red!15}0.5344&\cellcolor{red!15}0.3883\\
&3&-0.4530&0.0138&-0.6743\\ \hline
\multirow{2}{*}{23-Jun}&2&\cellcolor{red!15}-0.3357&\cellcolor{red!15}0.5199&\cellcolor{red!15}-0.1127\\
&3&-0.3987&-0.0296&-0.9632\\ \hline
\multirow{2}{*}{18-Jul}&2&-0.4434&0.5082&0.5518\\
&3&-0.5748&0.0622&-0.6809\\ \hline
\multirow{2}{*}{19-Jul}&2&-0.1517&0.7270&-1.1975\\
&3&-0.5790&0.0540&-1.5421\\ \hline
\multirow{2}{*}{03-Oct}&2&-0.4188&0.5411&1.7935\\
&3&-0.5648&0.0394&3.4696\\ \hline
\end{tabular}}
\end{table}

Table \ref{intervconf} contains the confidence/quantile intervals for the median coordinates of the groups of individuals defined by the observed star. .  For this case most of the medians are inside the quantile intervals (more than $70\%$ of the medians). The structure defined by the observed star is not as strong as the one defined the observation night. The quantile intervals from Tables \ref{mediancoord} and \ref{intervconf} correspond to a $96\%$ probability.  The stars can be identified in Table \ref{intervconf} with their Bayer designation (Greek letter followed by the abbreviation of the parent constellation's Latin name -  e.g.,$\zeta Vir$).
	
\begin{table}
\centering
{\footnotesize
\caption{Confidence intervals for individual groups defined by the observed star - dimensions 1 and 2} \label{intervconf}
\begin{tabular}[H]{|c|c|c|c|c|c|c|c|}
\hline
&\multicolumn{3}{c|}{$\alpha 2 Lib$}&\multirow{11}{*}{...}&\multicolumn{3}{c|}{$\zeta Vir$}\\ \cline{2-4} \cline{6-8}
&$Q^{2}_{50}$&$Q^{49}_{50}$&Median&&$Q^{2}_{50}$&$Q^{49}_{50}$&Median \\ 
\cline{1-4} \cline{6-8}
\color{gray}1&\color{gray}2&\color{gray}3&\color{gray}4& &\color{gray}233&\color{gray}234&\color{gray}235\\ \cline{1-4} \cline{6-8}
Dim.1&-1.37365&1.13447&\cellcolor{red!15}1.12740&&-0.88601&1.17979&1.95559\\ \cline{1-4} \cline{6-8}
Dim.2&-0.95837&0.81787&\cellcolor{red!15}-0.24940&&-0.98032&0.99207&\cellcolor{red!15}-0.10296\\ \cline{1-4} \cline{6-8}
Dim.3&-0.87625&0.4393&\cellcolor{red!15}-0.24627&&-0.8321&0.28378&-1.03948\\ \cline{1-4} \cline{6-8}
Dim.4&-0.96603&1.01562&\cellcolor{red!15}0.21106&&-0.7906&0.97161&\cellcolor{red!15}-0.07010\\ \cline{1-4} \cline{6-8}
Dim.5&-0.63005&0.38843&\cellcolor{red!15}-0.28810&&-0.9134&0.64677&\cellcolor{red!15}-0.16855\\ \cline{1-4} \cline{6-8}
Dim.6&-0.65236&0.85768&\cellcolor{red!15}-0.34874&&-0.57335&0.66941&\cellcolor{red!15}0.45769\\ \cline{1-4} \cline{6-8}
Dim.7&-0.69171&0.40592&0.71968&&-0.40571&0.32182&0.64569\\ \cline{1-4} \cline{6-8}
Dim.8&-0.37297&0.3473&\cellcolor{red!15}0.23813&&-0.41065&0.53837&\cellcolor{red!15}0.05889\\ \cline{1-4} \cline{6-8}
Dim.9&-0.55583&0.31309&\cellcolor{red!15}0.19781&&-0.37045&0.48317&\cellcolor{red!15}0.06883\\ \hline
\end{tabular}}
\end{table}

The linear correlation coefficient between the data set variables and the PCA axes, respectively, contains useful information. It reveals the contribution of one variable to the variability of the point cloud within a considered axis. The variables that are strongly correlated with the first four PCA dimensions are highlighted in Table \ref{coordvar}. All nine variables have the absolute value of the estimated linear correlation coefficient higher than 0.5 for at least two dimensions. Among them, the pressure $P$ and the temperature $T$ influence the variability of at least two PCA dimensions. The other way around, the axes $D_1$ and $D_2$ are strongly correlated with at least three variables, each.
	
\begin{table}
\centering
{\footnotesize
\caption{Variables coorrelations with the first four dimesnions} \label{coordvar}
\begin{tabular}[H]{|c|c|c|c|c|}
\hline
Variable&Dim.1&Dim.2&Dim.3&Dim.4\\ \hline
P&0.22&\cellcolor{green!15}0.61&\cellcolor{green!15}0.51&-0.35\\ \hline
T&\cellcolor{green!15}0.52&-0.06&-0.35&\cellcolor{green!15}-0.65\\ \hline
H&0.03&0.38&\cellcolor{green!15}0.83&0.04\\ \hline
rms1&\cellcolor{green!15}0.67&-0.44&0.39&0.03\\ \hline
img&\cellcolor{green!15}-0.77&-0.23&0.20&0.07\\ \hline
rms2&0.38&\cellcolor{green!15}-0.71&0.32&0.22\\ \hline
A&0.11&\cellcolor{green!15}0.55&-0.25&0.46\\ \hline
Z&0.37&0.08&-0.04&\cellcolor{green!15}0.56\\ \hline
velocity&\cellcolor{green!15}0.64&0.29&-0.22&0.16\\ \hline
\end{tabular}}
\end{table}
	
\begin{figure}[H]
\centering
\includegraphics[scale=0.41]{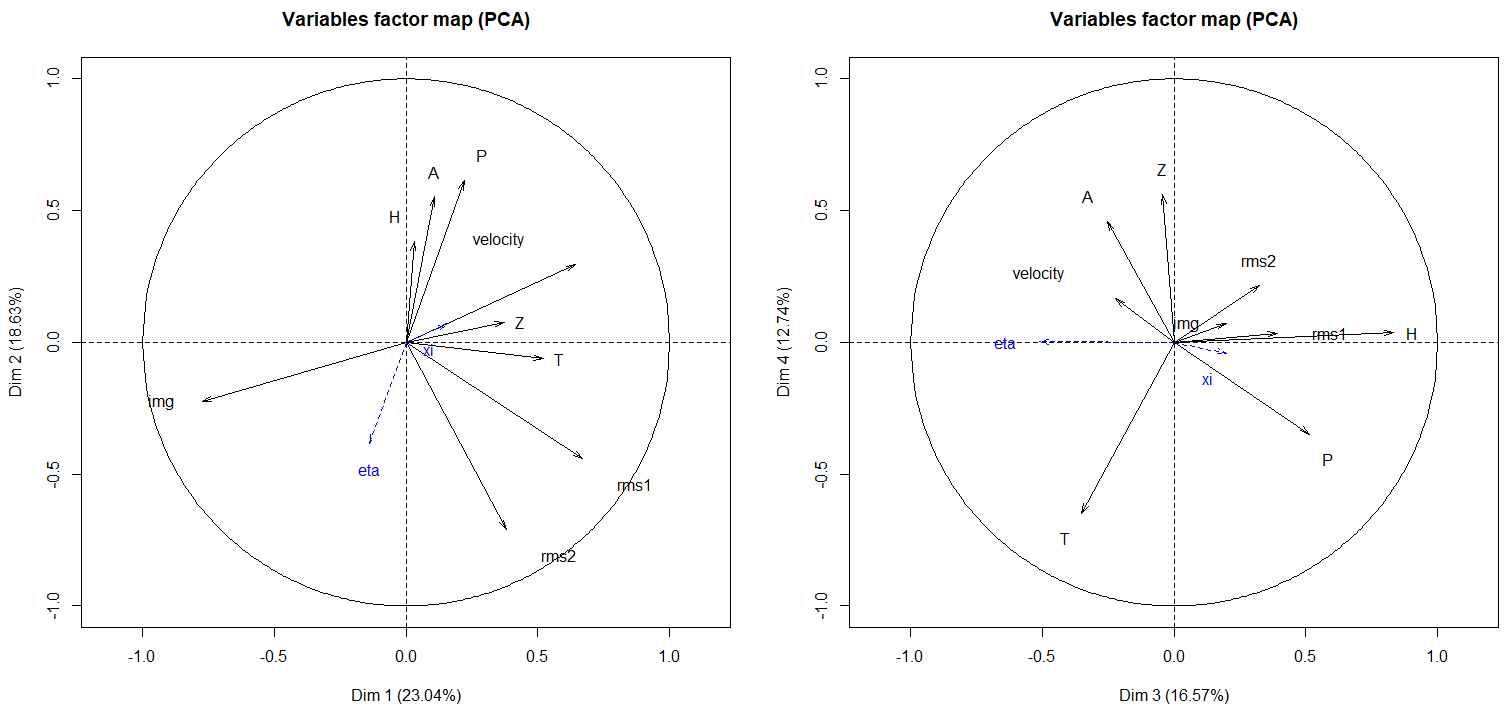}
\caption{Correlation circles for the first four PCA axes (1$^{st}$ and 2$^{nd}$ dimensions - right, 3$^{rd}$ and 4$^{th}$ dimensions - left, together with the two vertical deviation components.}
\label{varsuplimentare1}
\end{figure}

Since the correlation coefficient takes values in the interval $[-1,1]$, the correlation coefficients between the data set variables and the PCA axes can be represented on a hyper-sphere. Figure \ref{varsuplimentare1} shows correlation coefficients of the variables with the first four dimensions. The closer the correlation arrow is to the circle border, the stronger the linear dependence between the considered variable and axis. The $\eta$ and $\xi$ correlations were added to the plot. Hence from Figure \ref{varsuplimentare1} it can be concluded to what variables are the two vertical deviation components correlated with. The $\xi$ variable is strongly correlated with the variables "H", "A", "P", "velocity", "Z", "T" and "rms1" and the $\eta$ variable is correlated with the variables "img", "rms2", "T". This information may be important whenever linear modelling is considered. 

\section{MULTIPLE LINEAR REGRESSION MODELLING}
PCA results are an extremely useful exploratory tool. Nevertheless, in order to get more reliable inference the integration of this information should be done. This can be achieved through statistical modelling.In this section linear  modelling is used to check which of the variables in the data set that may explain the behaviour of the vertical deviation values.

The considered linear models are given by the regression equations:
\begin{equation} \label{eqregxi}
\xi_{i} = \beta_0^{\xi} + \sum_{j=1}^{p}X_{ij}\beta_{j}^{\xi} + \epsilon_{i}^{\xi}
\end{equation}
and 
\begin{equation} \label{eqregeta}
\eta_{i} = \beta_0^{\eta} + \sum_{j=1}^{p}X_{ij}\beta_{j}^{\eta} + \epsilon_{i}^{\eta}
\end{equation}
where $(X_{ij})$ are the elements of the design matrix given by Table~\ref{dataset}, the $\beta$s are the corresponding model parameters, $p$ is the order of the model and the $\epsilon$s are the corresponding errors, which here are assumed to be Gaussian i.i.d with their variances $\sigma^{2}_{\eta}$ and $\sigma^{2}_{\xi}$, respectively.

The data set presented in Table~\ref{dataset} contains a total of $510$ measures obtained during $12$ observation nights. Several measures are recorded during one night. The measures corresponding to one observation night are highly correlated with each other. In order to reduce the data correlation while taking into consideration all the available data, a bootstrap procedure is implemented, for solving the regression systems~\eqref{eqregxi} and~\eqref{eqregeta}.

The procedure is as follows. First, the design matrix size is fixed to $12 \times 9$, where $12$ is the number of nights and $9$ is the number of explanatory variables. Second, for each night a measure is chosen uniformly random, hence producing a random design matrix. Next, the regression model parameters are estimated. Finally, the last two steps of the procedure are repeated $100$ times, enabling the approximation of the probability distributions of the multiple linear regression outputs.  This was applied on the standardised data as it was used during the PCA analysis.

\begin{figure}[h]
\centering
\includegraphics[scale=0.41]{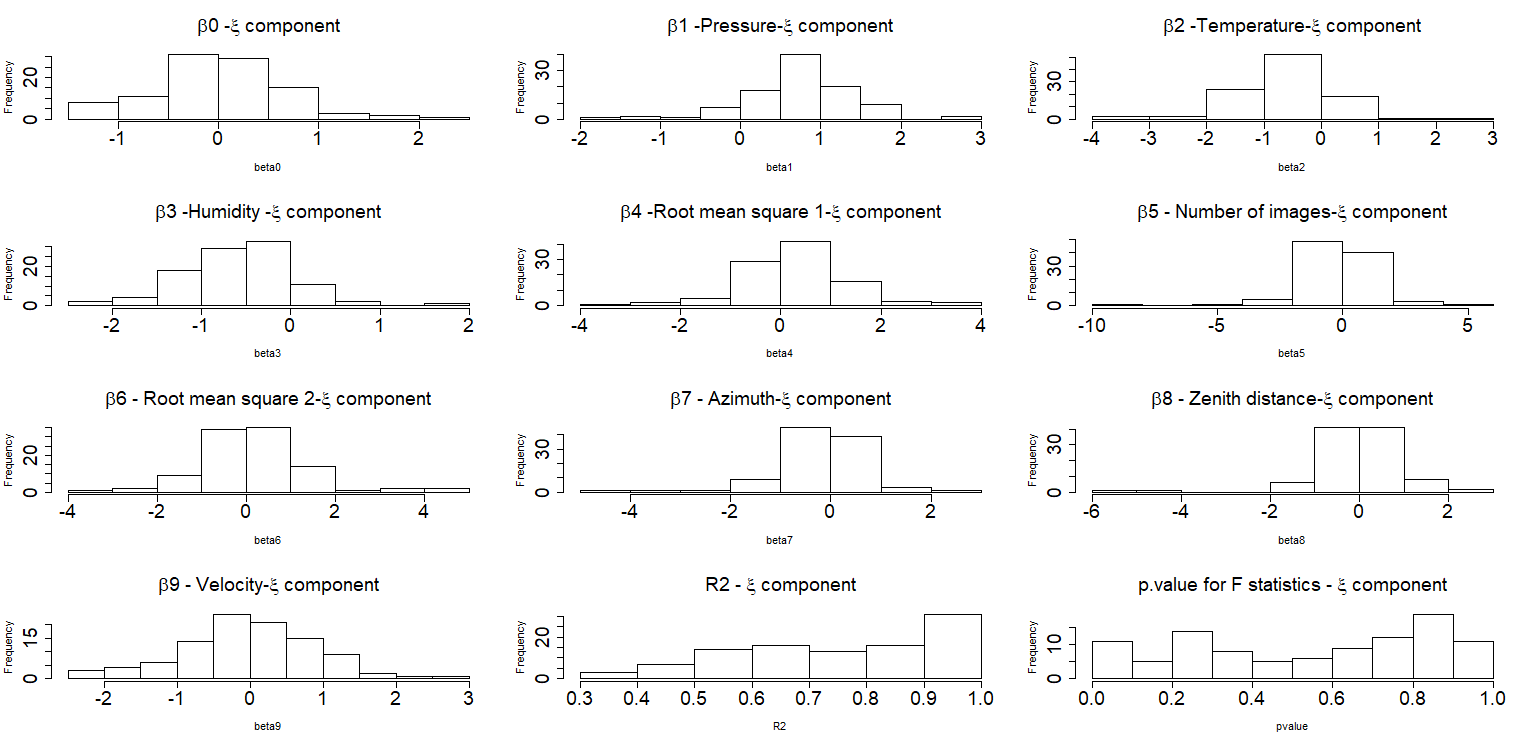}
\caption{Estimation distributions of the model~\eqref{eqregxi}.}
\label{histparamr2xi}
\end{figure}

Figure~\ref{histparamr2xi} shows the outputs of the method applied to the model~\eqref{eqregxi}. The model fits poorly the data. This is indicated by the $p-$values distribution of the Fisher test that does not reject the constant model. In agreement with this, the confidence intervals outlined by the model's coefficients distributions contain $0$ with high probability. Still, the histogram of the $R^2$ coefficient indicates that more than $50\%$ of the variation of the deviation angle may be explained by linear regression of the explanatory variables. The coefficient distributions corresponding to the pressure, the temperature and the humidity exhibit an asymmetric behaviour. This character may be interpreted as a particular weight that these variables may contribute within the variation of the deviation angle. Nevertheless, at this point, in order to be able to perform more reliable analysis more data is needed.

\begin{figure}[h!]
\centering
\includegraphics[scale=0.4]{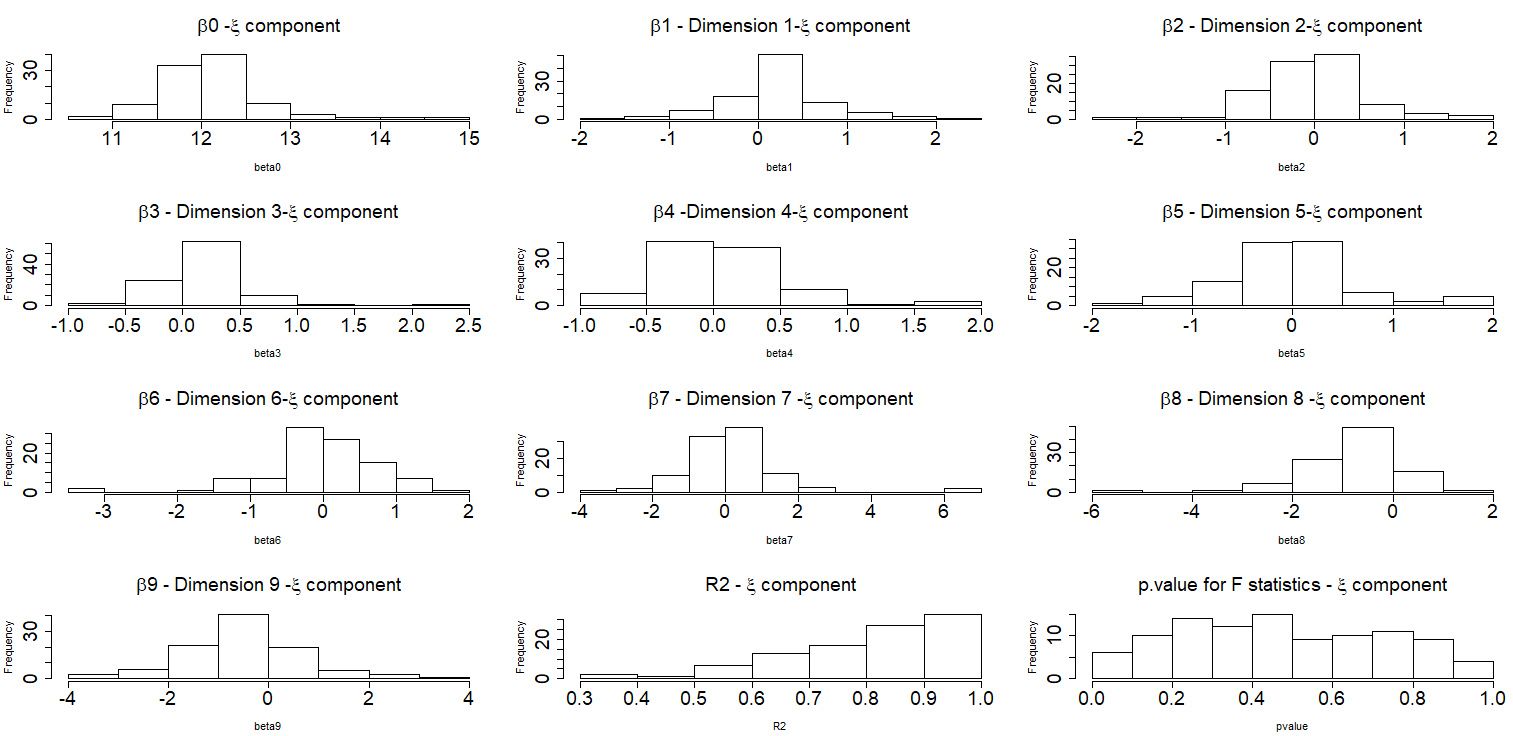}
\caption{Estimation distributions of the model~\eqref{eqregxi} applied to the PCA transformation of the design matrix.}
\label{histparamr2eta3}
\end{figure}

Following \cite{JacqFraix15}, the previous method was applied,  
while the design matrix for the models~\eqref{eqregxi} and~\eqref{eqregeta}, respectively, was given by the PCA coordinates associated to each element in the data set. The results are shown in Figure~\ref{histparamr2eta3}. The results analysis is similar to the preceding one. It appears that again, the model coefficients of the dimensions exhibiting asymmetric distributions are the ones that are the most correlated with variables such that the pressure, the temperature and the humidity.

\section{CONCLUSIONS}
This work presented the application of statistical based methodology for the analysis of the vertical deviation angle. The ingredients of the methodology were~: Principal Component Analysis and the Multiple Linear Regression models. In order to overcome the difficulties related to data correlation, while taking advantage by all the information provided, theses two classical techniques were embedded within a bootstrap procedure.

The methodology was applied on real data. The obtained results indicated that the pressure, the temperature and the humidity are variables that may influence the measure of the vertical deviation.

The preliminary results presented here are the starting point of an ongoing project. The campaign of collecting data should continue in order to improve the quality of the obtained results. Since the quantity and the quality of the data will increase recent statistical methods should be considered~\cite{HastEtAl09}.

\begin{acknowledgements}
 The first and the last author are extremely grateful to Dr. Octavian Badescu, for helping with the data collection and for the fruitful scientific discussions.

\vspace{2mm}

This work was supported by a grant of the Ministry of National Education and Scientific Research, RDI Program for Space Technology and Advanced Research - STAR, project number 513
\end{acknowledgements}

\makeatletter
\def\@biblabel#1{}
\makeatother

  \end{document}